# Turbo-like Iterative Multi-user Receiver Design for 5G Non-orthogonal Multiple Access

Xiangming Meng, Yiqun Wu, Chao Wang, and Yan Chen
Huawei Technologies, Co. Ltd.
Email: {mengxiangming1, wuyiqun, wangchao78, bigbird.chenyan}@huawei.com

*Abstract*—Non-orthogonal multiple access (NoMA) as an efficient way of radio resource sharing has been identified as a promising technology in 5G to help improving system capacity, user connectivity, and service latency in 5G communications. This paper provides a brief overview of the progress of NoMA transceiver study in 3GPP, with special focus on the design of turbo-like iterative multi-user (MU) receivers. There are various types of MU receivers depending on the combinations of MU detectors and interference cancellation (IC) schemes. Link-level simulations show that expectation propagation algorithm (EPA) with hybrid parallel interference cancellation (PIC) is a promising MU receiver, which can achieve fast convergence and similar performance as message passing algorithm (MPA) with much lower complexity.

I. INTRODUCTION

A non-orthogonal multiple access (NoMA) transmitter maps a stream of coded binary bits of a user (UE) to the available transmission resources by some user-specific operations to facilitate decoding of the superposed multi-user data at the receiver side with reasonable complexity. For the design of the user-specific operations, both power domain and code domain user separation schemes have been proposed from academic and industry. There are several overview literatures of the NoMA schemes and research progress for interested readers [1]-[4].

Note that though NoMA is beneficial for both uplink (multiple users to base station) and downlink (base station to multiple users), the focus of the current 3GPP study is in the UL. In this case, the NoMA receiver is equipped at the base station, which gives more design possibilities due to its strong processing capability compared with user equipment.

In this paper, we will briefly introduce the progress of NoMA study in 3GPP and then focus the discussion on one end of the NoMA design, i.e. the multi-user receiver. It is expected that the NoMA receiver design can show good BLER performance with fast convergence, and can be universally applied to all potential NoMA schemes and are friendly to real implementations.

A. *Progress of NoMA Transceiver Study in 3GPP*

The NoMA concept together with 15 different schemes were proposed as candidates for 3GPP Rel-14 NR (New Radio) study in early 2016. Most of the evaluations at that time focused on the mMTC scenario and positive conclusions were drawn from extensive simulations justifying the benefits of NoMA over orthogonal multiple access (OMA) in the capability of providing much larger connection density given the same system outage probability (measured in system packet drop rate) [5].

The study was suspended before making any recommendation on the transceiver design due to the very limited time budget in the overall NR study item. In Rel-15, a dedicated NoMA study item was approved and kicked off since Feb 2018. The focus of the study includes the NoMA transmitter side design in bit or symbol domain, NoMA receiver side design and complexity analysis, NoMA related procedures such as resource and signature allocation, HARQ feedbacks/combinations and link adaptation, and contention based grant-free NoMA with blind UE detection, as well as comprehensive link-level and system-level evaluations of NoMA in all three typical 5G scenarios, i.e., eMBB (enhanced mobile broadband), URLLC (ultra-reliable and low-latency communications), and mMTC (massive machine-type communications) with practical impairments and implementation constraints in each scenario, respectively.

B. *General Framework of Turbo-like Iterative Multi-user Receiver*

A general NoMA receiver has been agreed in the recent 3GPP meeting RAN1#92b [6]. As shown in Fig. 1, the high level diagram is adopted as the general block diagram of multi-user receiver for UL data transmissions. The algorithms for the detector block (for data) can be e.g. MMSE (minimum mean square error) [7], MF (matched filtering), ESE (elementary signal estimator) [8], MAP (maximum a posterior), MPA (message passing algorithm) [9], and EPA (expectation propagation algorithm) [10]. The interference cancellation (IC) can be hard or soft, and can be implemented in serial or parallel. Note that the IC block may consist of an input of the received signal for some types of IC implementations. The interference cancellation block may or may not be used. If not used, an input of interference estimation to the decoder may be required for some cases. The input to interference cancellation may come directly from the Detector for some cases

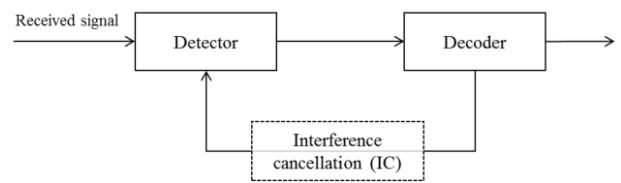

Fig. 1. A high-level block diagram of multi-user receiver[6].

In this paper, under the agreed high-level diagram in Fig. 1, we discuss several typical multi-user (MU) detectors, i.e., MPA, ESE, MMSE, and EPA. Moreover, different ways of interference cancellation that iterates information between the MU detector and decoder are also discussed. Implementation considerations together with link-level evaluations are used to analyze the different combinations of the MU detectors and IC schemes, from which the EPA hybrid PIC receiver is recommended for its fast convergence in BLER performance, universal application for all NoMA schemes, and friendliness in real implementations.

## II. MULTIUSER DETECTOR ALGORITHMS

In this section, some typical candidate MU detectors are described, including MPA, ESE, MMSE, and EPA.

### A. MPA

MPA (Message passing algorithm) [9] is an iterative MU detector with near-ML (maximum likelihood) detection performance. It passes messages that represent conditional probabilities back and forth between every FN (function node, representing resource element (RE)) and VN (variable node, representing data layer) edge in the factor graph of a NoMA scheme. To avoid confusion with the iteration between detector and decoder, the iteration inside MPA or other detectors is called inner-loop iteration, while the iteration between detector and decoder is called outer-loop iteration. In each inner-loop iteration, the messages between FNs and VNs are updated respectively. After a number of inner-loop iterations, the LLRs for the coded bits are calculated based on the current probabilities and then input to the channel decoder [11].

Its arithmetic complexity order is $O(M_P^{d_f})$ per inner-loop iteration, where $M_p$ denotes the number of the points on one RE constellation corresponding to a $log_2 M$-bit mapping. $M \geq M_p$, e.g., $M=8$, $M_p=4$; $M=16$, $M_p=9$; $M=64$, $M_p=16$. $d_f$ denotes the number of the (data) layers colliding over each RE. This complexity order can be further reduced by restricting the maximum number of layers: $d_s <= d_f$, from $O(M_P^{d_f})$ to $O(M_P^{d_s})$. This MPA with SIC is called SIC-MPA in [11].

### B. ESE

ESE (elementary signal estimator) [8] simply approximates the ISI (inter-user interference) plus Gaussian noise as Gaussian. Such a Gaussian approximation can be implemented in different ways for multiple receiving antennas: if a base-station performs matched filtering (MF) in the spatial domain, the approximation is a scalar Gaussian variable; if a base-station treats all receiving antennas jointly, the approximation is a joint Gaussian vector, i.e., multivariate Gaussian, which incurs high complexity.

Moreover, an ESE detector has to rely on the outer-loop iterations to achieve an acceptable detection performance and the convergence speed is slow. In case of high spectrum efficiency and high overloading, the number of outer-loop iterations may be too large for base station to reach short latency.

### C. MMSE

MMSE approximates the prior distribution of the signal as Gaussian whose mean and variance are computed from either soft LLRs fed back by the channel, or a Gaussian approximation with zero mean and variance scaled by the signal power (if the soft feedback is unavailable) [7]. A NoMA scheme with a SF (spreading factor) of $L$ can have two alternatives: a chip-by-chip MMSE that is performed on each RE independently, or a block-wise MMSE that is performed jointly on the $L$ REs. Matrix inversion dominates the MMSE complexity. A chip-by-chip MMSE needs to inverse $N_r$-by-$N_r$ complex-valued covariance matrix. A block-wise MMSE needs to inverse $N_r \times L$-by-$N_r \times L$ complex-valued covariance matrix. As a result, block-wise MMSE has much higher complexity, i.e., $\mathcal{O}(N_r^3 L^3)$ than chip-by-chip MMSE, i.e., $\mathcal{O}(N_r^3 L)$, especially when the spreading factor $L$ is large. While in base station implementations, higher order of matrix inversion could be less stable and harder to perform parallelization.

### D. EPA

EPA (expectation propagation algorithm) is one kind of well-known approximate Bayesian inference algorithm that has been widely used in machine learning [13][14]. It projects the target distribution of the transmitted symbols into a family of Gaussian distributions by iteratively matching the means and variances with the target distribution, which equivalently minimizes the Kullback-Leibler divergence between the target distribution and the approximate distribution [13].

EPA can be regarded as a type of Gaussian approximation to MPA but with consideration of the non-Gaussian nature of the transmitted symbols as well. It can also be viewed as an enhancement to ESE by iteratively refining the Gaussian approximation of the prior distribution. It has linear complexity with respect to $M$ ($M_p$ if low projection mapping is used) and $d_f$ ($d_s$ if SIC-EPA is used as SIC-MPA), while it provides nearly the same performance as MPA in most scenarios of interest [10]. The implementation of EPA can also employ the divider-and-conquer method and supports full parallelism.

For ease of comparison, we summarize various MU detectors discussion in Table 1.

**Table 1: Brief Summary of various MU Detectors**

| MU Detector | Basic Principle | Properties |
|---|---|---|
| MPA | Sum-product message passing performed on the factor graph of NoMA transmission | • Near ML detection performance<br>• Comparatively high complexity at high overload<br>• SIC-MPA as a low-complexity variant |
| ESE | Interference plus noise is approximated as Gaussian | • Comparatively low convergence rate at high overload and high SE |
| MMSE | The prior distribution is directly approximated as Gaussian | • Block-wise MMSE has much higher complexity than chip-by-chip MMSE |
| EPA | Gaussian approximation of MPA | • Fast convergence<br>• Nearly the same performance as MPA |

## III. INERFERENCE CANCELATION

In one dimension, the interference cancellation (IC) can be serial or parallel.

**SIC** (Serial IC, also known as successive IC) decodes only one user at a time, as shown in Fig. 2 (a). Although the order of SIC depends on SINR values to take advantage of near-far effect among users, such an ordered SIC may bring about a well-known error propagation. To overcome it, in enhanced SIC, the order of SIC is revised each time a UE is successfully decoded.

**PIC** (parallel IC) decodes all the active users simultaneously, thereby avoiding the order-related error propagation of SIC and improving the performance. In addition, it has low decoding latency due to higher parallelism. Note that one can also decode a subset of users at a time, which is called group PIC and may acquire additional operations for group selection.

In another dimension, IC can be hard or soft.

**Hard IC**: the channel decoder feedbacks the correctly decoded (i.e., passed cyclic redundancy check (CRC)) binary bits to the detector for the interference reconstruction. It feedbacks only correctly decoded data streams.

**Soft IC**: the channel decoder feedbacks the LLRs (log likelihood ratios) to the detector for the interference reconstruction, no matter whether the data stream can be correctly decoded or not. However, it may be a waste to use LLR values rather than hard IC for those successfully decoded codewords.

In practice, different combinations of SIC/PIC and Hard/Soft lead to different implementations. For example, the hard SIC shown in Fig. 2 (a) is a combination of SIC and hard IC, while the soft PIC in Fig. 2 (b) is a combination of PIC and soft IC. We propose a kind of hybrid PIC as shown in Fig. 2 (c),

**Hybrid PIC**: a combination of hard IC and soft IC in PIC. Specifically, for the successfully decoded (i.e., CRC passed) user or users, the interference is hard canceled; for the unsuccessfully decoded user or users, only soft LLRs are fed back by the decoder. Hybrid PIC can take advantages of both pure soft and pure hard IC schemes and achieves the best performance with low-complexity implementations.

## IV. EXPERIMENTAL RESULTS

In this section, link level simulations are performed to evaluate the turbo-like iterative MU receiver with different combinations of MU detectors and IC schemes.

For any type of IC scheme, the performance of the iterative MU receiver is closely related to the number of iterations between the channel decoder (backend) and the MU detector (frontend), also known as outer-loop (OL) iterations. To show the necessity of OL iterations, we set up CB-OFDMA [1] (contention-based OFDMA) and NLS [2] (Non-sparse linear Spreading) as two examples for transmitter processing. The channel model follows the TDL-A model [5] with delay spread 30ns and the moving speed is 3km/h. Details of simulation parameters are listed in Table II. As shown in Fig. 3, the block error rate (BLER) performance improves along with the increased number of outer-loop iterations until the performance is converged. If there is no outer-loop iterations, i.e., OL = 0, the performances of both CB- OFDMA and NLS are rather poor.

Moreover, the more multiplexed users/layers, the larger the required iterations to achieve convergence. As a result, outer-loop iterations are essential to obtain satisfactory performance.

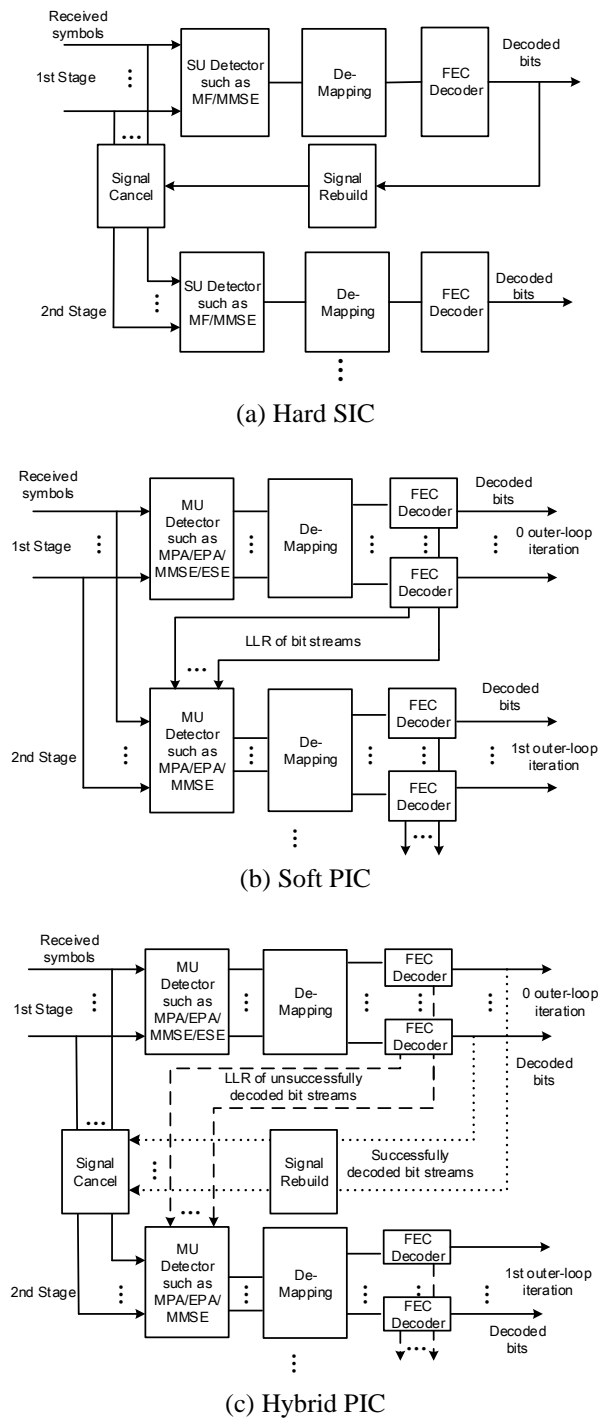

Fig. 2. Different IC schemes

---

[1] CB-OFDMA here refers to the contention based OFDMA where all the UEs fully share the same time and frequency resource also known as power-domain NOMA [4].

[2] NLS here means non-sparse linear spreading based NOMA schemes, which is a type of code-domain NOMA [18].

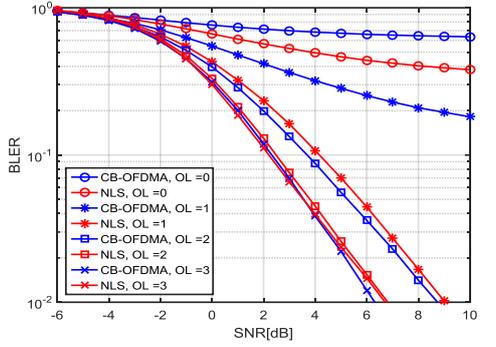

(a) 6UEs, 6RBs, 60bytes, 2Rx, OL 0,1,2,3.

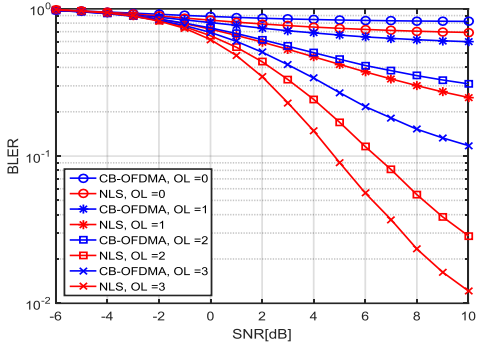

b) 8UEs, 6RBs, 60bytes, 2Rx, OL 0,1,2,3.

Fig. 3. BLER performances of CB-OFDMA and NLS with different number of outer-loop iterations.

**Table II: Simulation Setting.**

| Parameters | Values or assumptions |
|---|---|
| NoMA scheme | CB-OFDMA, NLS, SCMA |
| Carrier frequency | 700 MHz |
| Numerology | 14 OFDM symbols with 2 for DMRS |
| Transmission Bandwidth | 6 Resource Blocks (RBs) |
| Transport Block Size (TBS) | 60 bytes |
| Channel coding | NR Rel-15 LDPC |
| BS antenna configuration | 2 Rx |
| UE antenna configuration | 1Tx |
| Transmission mode | TM1 (refer to TS36.213) |
| Number of Multiplexed UEs | 6, 8, 10 |
| Propagation channel & UE velocity | TDL-A 30ns, 3km/h |
| Receivers | MMSE/ESE/MPA detector combined with different IC schemes |

We then evaluate and compare the performance of four different IC methods, namely hard SIC, enhanced SIC, soft PIC, and hybrid PIC. The MU detector is MMSE. Detailed simulation assumption is listed in Table II. For soft PIC and hybrid PIC, the number of outer-loop iterations is $OL = 3$. As shown in Fig. 4, hard SIC has the worst performance. By contrast, hybrid PIC can take advantages of both pure soft and pure hard IC schemes and achieves the best performance without much increase of complexity.

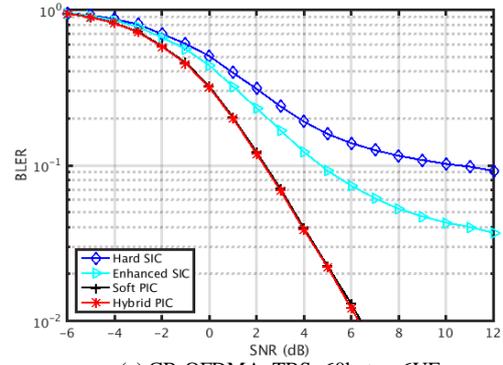

(a) CB-OFDMA, TBS=60bytes, 6UEs.

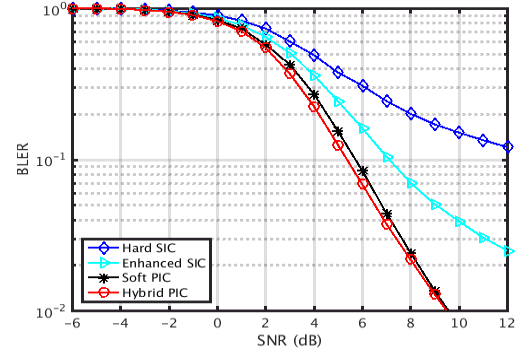

(b) NLS, TBS=60bytes, 10UEs.

Fig. 4. Performances comparison of different IC schemes

Next, under the hybrid PIC structure, we compare the performances of different MU detectors: ESE, MMSE, EPA, and MPA. MPA is a kind of near-ML detector and thus serves as the upper bound of the detectors. We consider CB-OFDM with 6UEs. Details of simulation parameters are listed in Table II. The number of outer-loop iterations is set to be $OL = 2$ and 3, respectively. As shown in Fig. 5, EPA can achieve nearly the same performance as MPA with $OL = 2$ and 3 while ESE and MMSE has apparent performance loss compared with MPA, especially with less number of outer-loop iterations. EPA converges much faster than ESE and MMSE, i.e., given the fixed number of outer-loop iterations, EPA can achieve much better performance then ESE and MMSE. This is quite appealing since fast convergence leads to low decoding latency which is quite important for 5G applications. Moreover, fast convergence of EPA also implies less times of channel decoding, which significantly reduces the receiver complexity.

Finally, given the receiver, the transmitter design can offer additional gains. To verify this, we compare the performances of different NoMA schemes: SCMA, CB-OFDMA, and NLS. The SCMA codebook used in the simulation is $M = 16$ point codebook proposed in [17]. The signatures of Non-sparse Linear spreading scheme is listed in Table A-2 of [18] with spreading length $L = 4$. Details of simulation parameters are listed in Table II. From previous discussion, EPA has much lower complexity compared with MPA and has better convergence performance compared with MMSE and ESE. Thus, in comparing different NoMA schemes, we chose the receiver to be EPA with hybrid PIC and the number of outer-loop iterations is set to be OL = 3. Fig. 6 shows that SCMA outperforms both CB-OFDMA and NLS, which implies that transmitter side symbol-level design is beneficial and the sparse feature can help improve the convergence and thus the BLER performance.

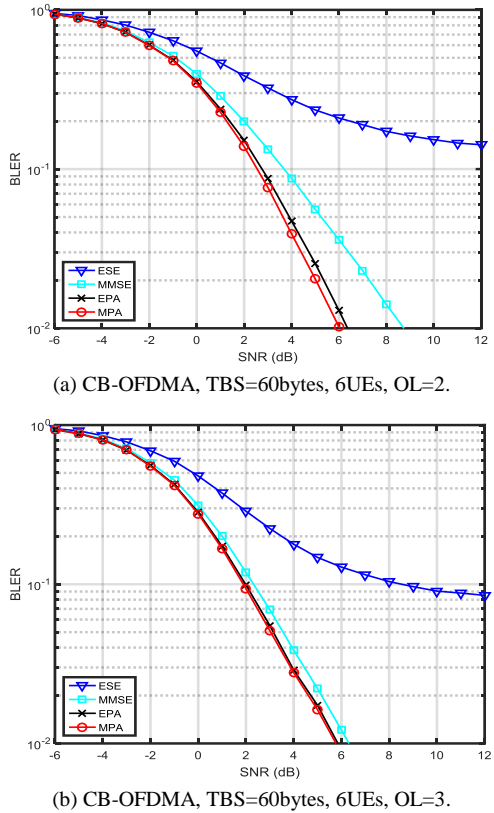

(a) CB-OFDMA, TBS=60bytes, 6UEs, OL=2.

(b) CB-OFDMA, TBS=60bytes, 6UEs, OL=3.

Fig. 5. BLER performance of MU detectors with hybrid PIC.

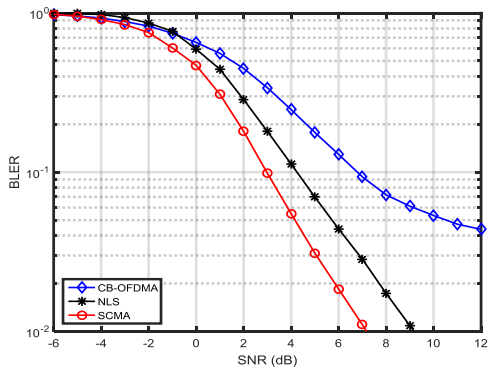

Fig. 6. Performances comparison of different NoMA schemes. TBS= 60bytes, 8UEs, EPA with hybrid PIC, OL = 3.

## V. CONCLUSION

In this paper, we introduce the recent progress of NoMA transceiver study in 3GPP and mainly focus on the general turbo-like iterative multi-user (MU) receiver design for 5G NoMA schemes. Under the general iterative structure, several candidate MU detectors, e.g., MPA, ESE, MMSE, and EPA, are summarized and compared. Moreover, different ways of interference cancellation that iterates information between the MU detector and decoder are also discussed. To take advantages of both pure soft and pure hard IC schemes, we propose hybrid PIC. Link-level simulations show that expectation propagation algorithm (EPA) with hybrid parallel interference cancellation (PIC) is a promising MU receiver, which can achieve fast convergence and similar performance as message passing algorithm (MPA) with much lower complexity.